# INSTALLATION PROGRESS AT THE PIP-II INJECTOR TEST AT FERMILAB*

C. Baffes[†], M. Alvarez, R. Andrews, A. Chen, J. Czajkowski, P. Derwent, J. Edelen, B. Hanna, B. Hartsell, K. Kendziora, D. Mitchell, L. Prost, V. Scarpine, A. Shemyakin, J. Steimel, T. Zuchnik, Fermi National Accelerator Laboratory, Batavia IL, USA
A. Edelen, Colorado State University, Fort Collins CO, USA

*Abstract*

A CW-compatible, pulsed H$^-$ superconducting linac "PIP-II" is being planned to upgrade Fermilab's injection complex. To validate the front-end concept, a test accelerator (The PIP-II Injector Test, formerly known as "PXIE") is under construction. The warm part of this accelerator comprises a 10 mA DC, 30 keV H$^-$ ion source, a 2 m-long Low Energy Beam Transport (LEBT), a 2.1 MeV Radio Frequency Quadrupole (RFQ) capable of operation in Continuous Wave (CW) mode, and a 10 m-long Medium Energy Beam Transport (MEBT). The paper will report on the installation of the RFQ and the first sections of the MEBT and related mechanical design considerations.

## PIP-II CONTEXT

Proton Improvement Plan II (PIP-II) [1,2] is a planned upgrade to the Fermilab accelerators complex that will enable the complex to deliver >1 MW of proton beam power on target to the Long Baseline Neutrino Facility (LBNF) [3], and provide a platform for future upgrades to multi-MW/high duty-factor operations. The existing normal-conducting linac will be replaced with a new superconducting RF linac injecting into the Booster Ring at 800 MeV. The PIP-II Injector Test (here abbreviated as PI-Test) [4] is a prototype of the front end of PIP-II, including the room-temperature accelerating section, a Half-Wave Resonator Cryomodule (HWR) [5] and a Single Spoke Resonator Cryomodule (SSR1). PI-Test is designed to accelerate 2 mA of CW beam to 25 MeV.

## PI-TEST STATUS AND PLANS

As of this writing, the ion source, LEBT [6] and RFQ have been fully installed. The RFQ has been conditioned with CW RF up to 100 kW input power, and commissioned with pulsed beam up to a 5 mA, 5 ms pulse at 10 Hz [7]. The MEBT [8] is evolving in stages designed to first verify RFQ requirements. The first stage (Fig. 1) included permanent components (two doublet/corrector magnet packages, designed and fabricated by Bhabha Atomic Research Centre (BARC), beam scrapers, and a bunching cavity), and instrumentation to characterize the beam coming out of the RFQ. Instrumentation included toroids for transmission measurements, a time-of-flight movable beam position monitor (TOF) to determine the beam energy, and a fast Faraday cup (FFC) for longitudinal bunch structure measurement. The transmission of the RFQ was verified to be ~98% and the energy to be 2.11 MeV ± 1% [7], both within specification.

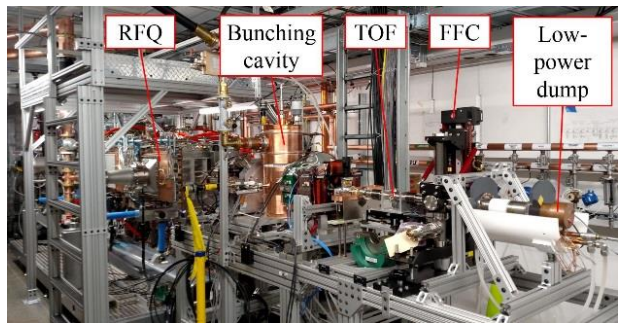

Figure 1: RFQ exit and MEBT in longitudinal diagnostic configuration for initial beam commissioning of RFQ

During a shutdown in September 2016 the MEBT was reconfigured to accommodate CW beam from the RFQ. A prototype absorber [9] was installed to investigate the potential for blistering from H$^-$ beam, and a long drift space and high-power beam dump [10] were installed to allow working with current up to 10 mA CW (Figs. 2, 4) and large beam size at the dump. As of this writing, this configuration is being commissioned in pulsed mode.

### Radiation from 2.1 MeV H- Beam

Measurable radiation, dominated by high energy photons (γ), has been observed from the 2.1 MeV pulsed beam incident on a Faraday cup made of copper, scrapers made of molybdenum alloy TZM and the nickel dump, and interpreted via simulation in the MARS program. Scaling these simulations/measurements up to 100% duty factor, predicted radiation levels in some areas of the enclosure are too high to permit beam-on access (e.g. ~100 mrem/hr adjacent to the dump). As a result, personnel interlocks are required for work with CW beam.

___________________________________________
* This work was supported by the DOE contract No.DEAC02-07CH11359 to the Fermi Research Alliance LLC.
† cbaffes@fnal.gov

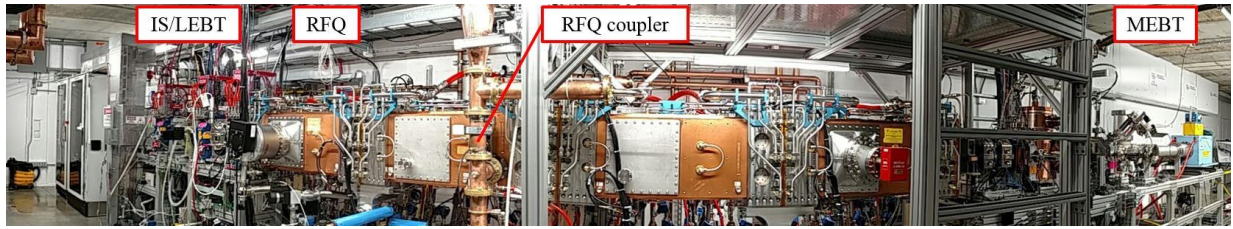

Figure 2: PI-test accelerator in RFQ CW-commissioning configuration

## Next steps for the MEBT

After demonstration of CW beam operation, the MEBT will be reconfigured to accommodate an Allison-type [11] emittance scanner. After measuring the transverse emittance, the MEBT will be extended to include 4 triplets (provided by BARC) and prototypes of the fast kickers designed to provide bunch-by-bunch chopping at 162.5 MHz. The final stage of the MEBT build will incorporate the Ultra-High Vacuum (UHV) low-particulate section that interfaces to the HWR cryomodule.

## MECHANICAL IMPLEMENTATION

### Vacuum Systems

The high gas load from the ion source and from H- recombination motivates the use of turbo pumps as the primary pumps in the current portion of the machine. The UHV section of the MEBT will be exclusively ion-pumped, and separated from the upstream MEBT by a differential pumping orifice. The heavy use of turbo pumps and their backing scroll pumps creates several locations where mechanical failure is possible. In order to ensure reliability and machine safety, a preventative maintenance program has been initiated for the pumps. Failure of scroll pumps is of greatest concern. Hence, the vacuum volume is protected by an anti-suckback valve in each roughing line which is directly controlled by a local convection gauge to eliminate external controller processing time.

### RFQ Cooling

The RFQ does not include any movable tuners. Tuning is accomplished by modulating the water temperature in two circuits – one which cools the RFQ vanes and another which cools the RFQ outer walls. This approach requires a cooling system with short response time and excellent stability, on the order of ±0.1°C. This is accomplished with a mixing architecture (Fig. 3). While warm water is recirculated through the RFQ, a variable amount of cold water is injected through a variable control valve to achieve the desired supply temperature. A key concern is the temperature uniformity of the cooling water. This is accomplished with a multiple-branch distribution system to each of the four modules of the RFQ, and extensive manual valving to balance the flow. Also, the flow through the RFQ is large (~200 gpm total) to minimize temperature rise and associated thermal gradients. The initial implementation for resonance control provides proportional/integral control of both vane and wall cooling circuits [12]. A model predictive control implementation is in development [13, 14].

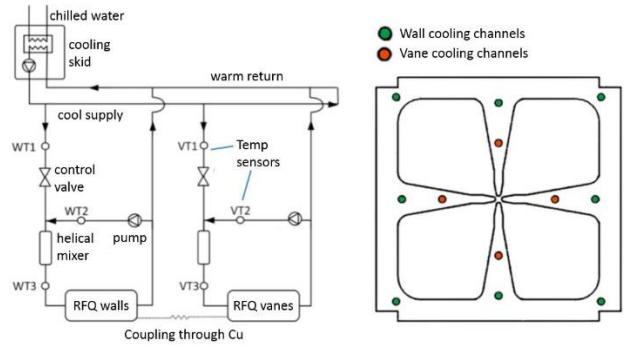

Figure 3: RFQ cooling system conceptual schematic

### MEBT Modular Design

During RFQ and MEBT commissioning, frequent reconfiguration is planned to accommodate temporary instrumentation and incremental expansion of the MEBT line. As such, an effort was made to keep the design modular at both the component level (i.e. individual pieces of instrumentation) and at the subassembly level (i.e. rafts or girders containing several doublet or triplet assemblies and interstitial beamline components).

Structures are built around a commercially-available T-slot extruded aluminum. This approach has proven to be both economical and flexible, and permits for easy accommodation of unforeseen structure, strain relief, etc.

Several components with the potential to be removed for repair are mounted on cone/"v"/flat kinematic mounts, such that alignment can be recovered without an extensive adjustment process.

### CAD Practices

At Fermilab, Computer Aided Design (CAD) is executed in the NX design program, with product data management through Teamcenter. Compared to most accelerators, the PI-Test accelerator is of modest size and scale (~40 m total length when complete). However, given the complexity of the components, this is already enough to create processing difficulties when running CAD on conventional computers. A standard approach used in the past is to break an accelerator up into small pieces (generally <10 m) with detailed models, and handle higher-level assemblies in a separate, simplified model. This has the disadvantage of breaking the bill of material traceability, and creates the strong possibility of discrepancy between models. Model simplification methods are built into the software (e.g. jt-based drawing view options), but as of this writing these methods do not allow for precise measurement or dimensioning of components on top-level assembly drawings.

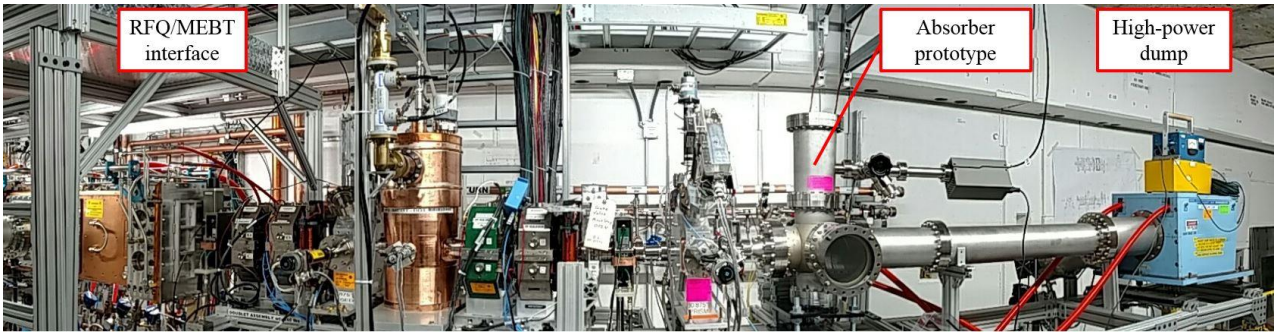

Figure 4: MEBT in RFQ CW commissioning configuration

In order to address these shortcomings, modeling practices have been developed that create simplified geometry at the device level (e.g. a triplet assembly or single piece of instrumentation). The simplified geometry resides as an arrangement (in NX parlance) within the device assembly. These simple arrangements are called by default at higher levels of assembly, and allow efficient rendering of top-level models and drawings, while maintaining correct bill of material structure and avoiding the need for parallel and possibly conflicting models. In any case where more detailed representations are needed (i.e. interface drawings at physical interfaces between devices or systems), the detailed arrangements may be loaded. Compliance with these practices is maintained through the use of a standard quality assurance checklist that is validated at the time of model/drawing release. These methods are being developed and refined in PI-Test so that a mature implementation is ready for PIP-II.

*Passive Vibration Isolation*

The cryomodules, especially SSR1, are sensitive to vibration and microphonics due to their narrow bandwidth [15]. With this in mind, PI-test is being designed from an early stage to minimize the generation and transmission of vibration to the cryomodules. The hardware described herein is several meters away from the SSR1, so passive vibration control is implemented based on non-analytical best-practice guidelines. All rotating equipment is soft-mounted on vibration mounts. Compliant couplings are used to minimize the transmission of vibration through cooling water plumbing (Fig. 5).

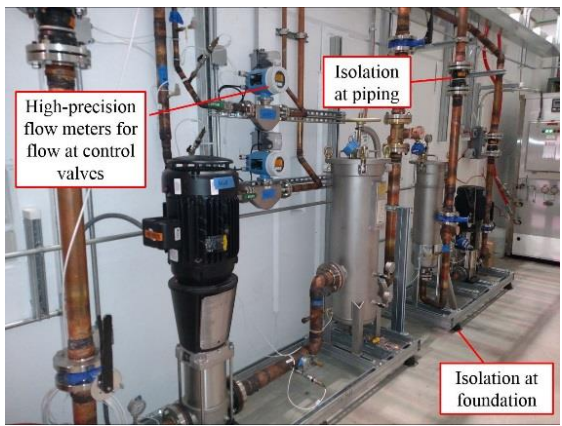

Figure 5: Vibration isolation at RFQ cooling skids

Vibrations within the facility, which is shared with other SRF projects, are intermittently monitored to identify and mitigate vibration sources. Components mounted within ~2m of the cryomodules will be subject to more rigorous characterization of vibration transfer function to the cavities.

## ACKNOWLEDGMENT

The authors wish to acknowledge the efforts of partner labs who provided key deliverables; the RFQ team at Lawrence Berkeley Lab - D. Li, A. DeMello, M. Hoff, A. Lambert, T. Luo, J. Staples, S. Virostek and their colleagues; and the MEBT magnet team at BARC - S. Malhotra, V. Teotia, K. Singh, J. Itteera and their colleagues.

The authors would also like to express gratitude to the large team of designers, engineers, scientists, and technical staff who executed the installation and commissioning of PI-Test.